# Particle manipulation beyond the diffraction limit using structured super-oscillating light beams

[1]Brijesh Kumar Singh*†, [2]Harel Nagar†, [2]Yael Roichman, and [3]Ady Arie

[1,3]School of Electrical Engineering, Fleischman Faculty of Engineering, Tel-Aviv University, Israel

[1]Department of Physics, Central University of Rajasthan, Ajmer, India

[2]Raymond and Beverly Sackler School of Chemistry, Tel-Aviv University, Israel

†These authors contributed equally to this work

*Corresponding author: brijeshsingh831@gmail.com

**The diffraction limited resolution of light focused by a lens was derived in 1873 by Ernst Abbe. Later in 1952, a method to reach sub-diffraction light spots was proposed by modulating the wavefront of the focused beam. In a related development, super-oscillating functions, i.e. band limited functions that locally oscillate faster than their highest Fourier component, were introduced and experimentally applied for super-resolution microscopy. Up till now, only simple Gaussian-like sub-diffraction spots were used. Here we show that the amplitude and phase profile of these sub-diffraction spots can be arbitrarily controlled. In particular we utilize Hermite-Gauss, Laguerre-Gauss and Airy functions to structure super-oscillating beams with sub-diffraction lobes. These structured beams are then used for high resolution trapping and manipulation of nanometer-sized particles. The trapping potential provides unprecedented localization accuracy and stiffness, significantly exceeding those provided by standard diffraction limited beams.**

**INTRODUCTION**

Gaussian beams, which are characterized by a transverse Gaussian intensity profile, are the most commonly used beams in laser optics. However, there are many other types of structured beams[1], representing different solutions of the underlying wave equation for light in free-space – the Helmholtz equation – that maintain, in general, their shape during propagation. Such beams include the Hermite-Gauss (HG) beams, having multiple intensity lobes, Laguerre-Gauss vortex beams that are characterized by helical phase and carry orbital angular momentum (OAM)[2,3], self-accelerating Airy beams that propagate along parabolic caustic trajectories[4] and more[5]. There are numerous applications that utilize these structured beams, for example, their linear momentum applies a reaction force that can trap particles, whereas the angular momentum, e.g. of vortex beams, can be used for rotating them[1, 6-10]. Vortex beams are also useful for depleting fluorescent dye molecules in Stimulated Emission Depletion (STED) microscopy[11], while Airy beams can be used for optically-mediated particle clearing[12] and light sheet microscopy[13]. Until recently, all the realizations of structured beams were limited in scale by the standard diffraction limit $\lambda/2NA$ (where $\lambda$ is the optical wavelength and NA is the lens numerical aperture). This limit originates from the beam size for a lens illuminated by a plane wave, and is also identical to the Abbe resolution limit[14]. However, when the lens is not homogenously illuminated, smaller features can be obtained at the focal plane. Already, in 1872 J.W. Strutt, also known as Lord Rayleigh has demonstrated that a smaller spot of 0.36λ/NA is achievable by modulating the input beam using an annular aperture[15]. Later G. Toraldo di Francia[16] proposed the concept of super-directive antennas could be applied to optical instrument to achieve the much narrower focal spot than the Abbe limit. By following the concepts that were developed for weak quantum mechanical measurements[17], Michael Berry introduced the concept of super-oscillations for the band limited functions that locally oscillate faster than their highest Fourier component[18]. The phenomenon of super-oscillations, exploited in the form of super oscillatory lens in optical microscopy, was proposed as a method to achieve super-resolution imaging[19,20]. In this case, since the lens generates a band-limited intensity pattern with a bandwidth of $2NA/\lambda$, a light field locally oscillates faster than that is super-oscillatory. Indeed, super-oscillating (SO) beams with sub-diffraction limit features (but without any internal structure) have been demonstrated by modulating the lens pupil[21,22] and by superposition of Bessel beams[23-25]. A question that arises is whether it is possible to structure these sub-wavelength light spots. If so, exciting new possibilities will emerge for particle manipulation[26,27] and super-resolution microscopy beyond the diffraction limit.

In this paper we present a systematic approach for structuring super-oscillating beams. We note that a different approach was proposed in[24] but was not explored experimentally up till now. Here we demonstrate our method by realizing SO beams with features that are smaller than half of the optical wavelength and having either multiple lobes, or a helical phase-front, or an Airy-like shape. In order to demonstrate the usefulness of these beams, we utilize them to trap a single nano-particle with unprecedented localization and stiffness using an SO-Gaussian beam with a single central lobe. In addition, we use multi-lobe SO-Gaussian beams to trap multiple nano-particles, SO-vortex beams to rotate particles clockwise and counter-clockwise at different angular velocities.

**MATERIALS AND METHODS**

**Mask Realization**

To introduce our approach to the realization of structured SO beams, let us first consider the case in which at the pupil plane of the lens an annular phase mask $\psi_{mask}(r)$ is inserted, consisting of two regions with opposite phase. The mask is expressed by[28,29]

$$\psi_{mask}(r) \propto \begin{cases} -1 & r \leq r_\pi \\ 1 & r_\pi \leq r \leq r_{max} \end{cases}, \qquad (1)$$

where $0 < r_\pi < r_{max}$, $r_{max} = D/2$, $D$ is the diameter of the mask aperture at the pupil plane and $r$ is the radial co-ordinate. Plane wave illumination of this binary phase mask will generate the Fourier transform $FT\{\psi_{mask}(r)\}$ at the focal plane of the lens. This generated beam is a super-oscillatory beam, having a central lobe whose size is determined by $r_\pi$, followed by a peripheral ring of light. To experimentally realize the super-oscillatory structured beams we implemented this binary mask into an off-axis computer generated hologram[30]. The carrier wave number $k_c$ of the hologram determines the separation of the unwanted zero order and the target SO beams in first order, using the following phase modulation pattern:

$$\psi_{holo}(r,x) \propto \begin{cases} \exp\{i(k_c x + \pi + \phi_{obj}(x,y))\} & r \leq r_\pi \\ \exp\{i(k_c x + \phi_{obj}(x,y))\} & r_\pi \leq r \leq r_{max} \end{cases}, \quad (2)$$

where $\phi_{obj}(x,y)$ is the phase function of the Fourier-transform of the desired super-oscillatory structured beams. The advantage of this method is that both the amplitude and phase of the wave functions can be realized in the first order of the diffraction pattern by modulating only the phase of the beam, hence it can be easily realized with a spatial light modulator. By changing $r_\pi$ the focal spot size of the SO beams can be made arbitrarily small, although the spot intensity is simultaneously reduced. In the supplemental material section we provide a numerical comparison of the local frequencies of an SO beam and a normal Bessel beam that is obtained from a circular aperture. We also show the variation of focal spot size of SO beam as a function of $r_\pi$.

**Experimental details**

We implement Eq. (2) by encoding it onto a phase only spatial light modulator (SLM). The SLM has 800 x 600 pixels with pixel pitch $20 \mu m$ used in the reflective mode. A spatially filtered and collimated linearly polarized Gaussian laser beam (Coherent, Verdi 6) at $\lambda = 532 nm$ is incident on the SLM. The reflected wave-fronts from the SLM, with the help of the telescopic lenses and dichroic mirror reached to the back aperture of 100X oil-immersion microscope objective (MO) lens with numerical aperture $NA = 1.4$ (Olympus IX-71 microscope). In the focal volume of MO, the desired profiles of the structured SO beams having the sub-diffraction focal spots are realized, as shown in Fig.1a. Different experimental realizations of three dimensional (3D) intensity distribution of the structured SO beams calculated according to Eq. (2) are presented in Fig. 1(b-e) using the volumetric imaging technique[31] i.e. reconstruction from 2D sections taken by imaging the beam as reflected from a mirror in the image plane. We characterize the size of our diffraction limited beam from the full width half maximum (FWHM) of its reflection from a mirror placed in the sample plane. The parameters of SO-Gaussian beam, SO-vortex beam, SO-HG beam and SO-Airy beam used in Eq. (2) are provided in the table. The phase mask, intensity distribution in the focal plane, and their comparison to the equivalent normal structured beams is presented in Fig.2. Since the effective numerical aperture in our case is $NA_{eff}=0.55<0.7$ the effect of input polarization of the beam on the focal plane intensity profile can be neglected[32] and so we used the scalar diffraction theory[33] to simulate the intensity profile in the focal plane. The Hermite-Gauss beam can also be realized by multiplying the exponential phase function $\exp(i(k_c x+\phi_{obj}))$ with the Hermite polynomial, as done in Fig.2d1. Clearly, sub-diffraction features can be obtained for the structured SO beams as well as for the non-structured SO beam, as will be demonstrated in the next section.

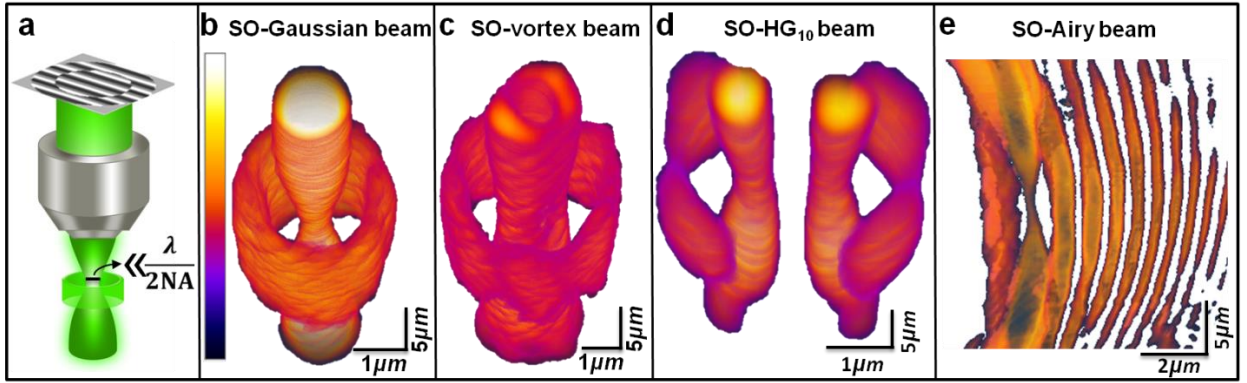

**Figure 1.** Super oscillating (SO) beams generation. **a** A schematic of the optical setup; optical wavefronts are modulated by the phase mask and focused through a microscopic objective lens to generate SO beams. **b-e** Three dimensional (3d) volumetric profiles of the generated SO-structured beams - SO-Gaussian beam (SO-GB), SO-vortex beam with $l=1$, SO-HG$_{10}$ beam, and SO-Airy beam, respectively. Beams were created with relatively large features to better image their structure. The threshold intensity levels for the contour plots in Figs.1 (b-e) are 0.43, 0.34, 0.36 and 0.31 of the maximum intensity, respectively.

| Functions | SO-Gaussian beam | SO-vortex beam | SO-Hermite-Gauss (HG) beam | SO-Airy beam |
|---|---|---|---|---|
| $\phi_{obj}$ | 0 | $l\theta$; $l = \pm 1, \pm 2...$ | $phase\left(H_m\left(\sqrt{2}\,x/w_o\right)H_n\left(\sqrt{2}\,y/w_o\right)\right)$; $m, n = 0, 1, 2...; w_o = 1mm$ | $(x^3 + y^3)/a^3$; $a = 4.14\mu m$ |

**RESULTS AND DISCUSSION**

We start by realizing a sub-diffraction limited laser spot based on a Gaussian SO beam. The theoretical beam waist of the focal spot for a Gaussian beam illuminating the entire lens aperture is calculated by $d = 1.27\lambda/2NA$ and its FWHM is $\approx 0.38\lambda/NA \approx 145nm$, but since in our set up we are using a Gaussian beam whose width is smaller than the lens aperture, the effective numerical aperture is $NA_{eff} = 0.55$. The measured $FWHM = (427 \pm 58)nm$ agrees well with theoretical expectation of $0.38\lambda/NA_{eff} = 368nm$, (see Fig. 2a$_5$). With the same illumination, the measured FWHM of the central lobe $= (163 \pm 58)nm \approx 0.31\lambda$ of a SO-Gaussian beam (generated with $\phi_{obj} = 0$, for $r_\pi = 0.65r_{max}$) is a factor of $\sim 2.6$ smaller (see Fig. 2a$_3$). Next we show the ability of this method to generate sub-diffraction "doughnut" shaped SO-vortex beams by substituting $\phi_{obj} = l\theta$ in Eq. (2). Owing to the helical phase term, the central lobe of this beam carries orbital angular momentum with topological charge $l = 1$. The measured dark core size with $r_\pi = 0.70r_{max}$ is $\approx (239 \pm 58)nm$, more than a factor of 1.3 smaller than the achieved core with a diffraction limited beam (compare Fig. 2b$_3$ and Fig. 2b$_5$). We also generated SO-vortex beam with a larger topological charge $l = 3$. For $r_\pi = 0.72r_{max}$ the dark core size is $(790 \pm 58)nm$, significantly smaller than the $(947 \pm 58)nm$ dark core size of the corresponding diffraction limited beam. Next, we generated self-similar super-oscillating HG$_{10}$ beams, having two lobes with identical size $\approx (173 \pm 58) nm$ for $r_\pi = 0.72 r_{max}$, where each lobe is significantly smaller than the diffraction limited spot size, see Figs. 2d$_3$ and 2d$_5$. The beam waist of SO-HG function is $=1mm$ in the phase mask, displayed on the SLM. Finally, we generated the SO Airy-like beam where for $r_\pi = 0.30 r_{max}$, the FWHM of the small lobe that is located to the right of the main lobe of the beam is $\approx (269 \pm 58) nm$. Its size is approximately only a quarter of the size of the main lobe of a normal Airy beam $\approx (1024 \pm 58) nm$ as shown in Fig.2e. Similar narrowing of the Airy lobe were observed recently under an appropriate transverse compression of their spatial spectra[34,35] and by superposition of Airy modes[36].

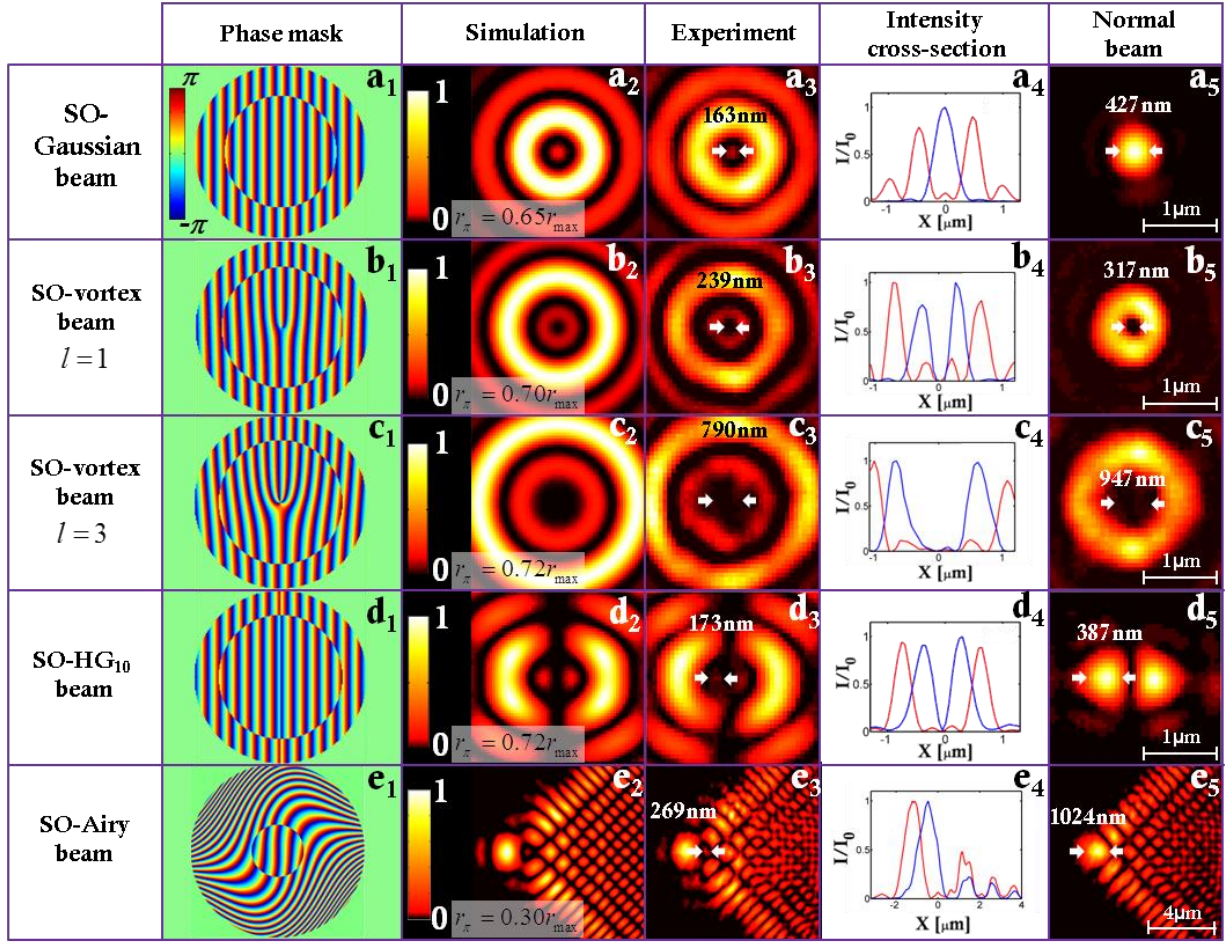

**Figure 2.** Gallery of generated SO beams: Gaussian beam, vortex beam of charge $l=1$ and $l=3$, $HG_{10}$ beam and Airy beam. All the SO phase masks, their corresponding simulation, and their experimental realizations are given in the first, second and third columns respectively, while the experimental realizations of their counterpart normal beams are given in the last column. One dimensional intensity plots of SO beams and their equivalent normal beams used to measure the FWHM are given in the fourth column in red and blue colors, respectively. Each beam profile is normalized to its own maximum.

Having established the superior focusing of SO beams, we are interested in their application to improving optical trapping. First, we demonstrate that optical trapping is possible with structured SO beams such as a SO-Gaussian beam, a SO-vortex, and SO-HG beams, as shown in Fig. 3 (see Supplemental movies 1-5). For this application we used slightly larger lobes, although they are still beyond the standard diffraction limit. If we use the same beam profiles as given in Fig.2, the central and first side lobes are too close to trap 500nm size particle at the central lobe of the SO beams and therefore the particle jumps from the central lobe to the side lobes. Hence, we chose to use slightly larger lobes where there is a sufficient gap between the central and side lobes, thereby enabling stable trapping of particles at the central lobe. The optical trapping force of dielectric particles is proportional to the gradient of the trapping laser intensity[37]. The localization quality is a function both of the focal spot size and the trapping force. It is therefore not straightforward to assume that a sub-diffraction limited spot would result in stronger trapping or better localization. To characterize the properties of SO-Gaussian beams as optical traps we compare the trapping of a single $500nm$ polystyrene bead by either a diffraction limited Gaussian beam (width $(450\pm58)nm$) or a SO-Gaussian beam (width $(188\pm58)nm, r_\pi = 0.67 r_{max}, D = 2.1mm$). Suspension of polystyrene beads, $500nm$ in diameter (Invitrogen lots #1173396), were placed between glass slide ($1.1mm$ in thickness) and cover slip ($0.15mm$ in thickness) to form an approximately $20\mu m$ thick sample. We use dilute suspensions to control the number of trapped particles, and laser intensity to control the strength of trapping. In this way we manipulated the particles to sit either in the central spot of a SO-Gaussian beam or the ring surrounding it against the slide. We use conventional video microscopy[38] to extract the trapped particles' trajectories from which we derive the two dimensional (2d) probability distribution of their position (see Figs. 3 b and c), and the corresponding trapping stiffness[39]. The trapping results presented below correspond to lateral (2D) trapping balancing radiation pressure against a glass wall. Although the maximal intensity of the central lobe of the SO-Gaussian beam is 7 times lower than that of the Gaussian beam, it is clear from Figs. 3b and 3c that the SO-Gaussian beam confines the particle significantly better (Supplemental movie 1). We performed single polystyrene bead trapping experiments for 11 different beads and observed the average standard deviation of positions indicates nearly 3-fold improvement in localization, $\sigma_{SO} \approx 14.9 nm$ for the SO-Gaussian beam versus $\sigma_{GB} \approx 44.6 nm$ for the Gaussian beam. We plot the histogram of the stiffness ratio between SO and normal Gaussian beam corresponding to the different trapping measurement of single polystyrene bead. The average localization improvement translates to a 9.4-fold improvement (see supplementary for details and Fig. 8) in the trapping stiffness from $K_{GB} \approx 2.1 pN/\mu m$ to $K_{SO} \approx 18.5 pN/\mu m$. Laser power arriving at the optical trap was 1mW and 0.025mW for Gaussian and SO-Gaussian beams traps, respectively. In terms of trapping efficiency (stiffness /mW), the central lobe of the SO beam shows more efficient trapping by the factor of 350 times than the normal Gaussian beam. This huge enhancement in the trapping efficiency of SO beam is the result of the big difference between the power of central SO beam and normal beam. The relatively large variation in the trapping stiffness of the SO beam is due to its high sensitivity both to the axial distance from the focal plane and the alignment of the system (for more details see supplementary). We note that 200*nm* polystyrene beads were also trapped using the SO-beam (see Fig.6a in Supplemental Material section), but we did not use them for analyzing the trapping stiffness, owing to their smaller size and resulting difficulty in precisely quantifying their movement.

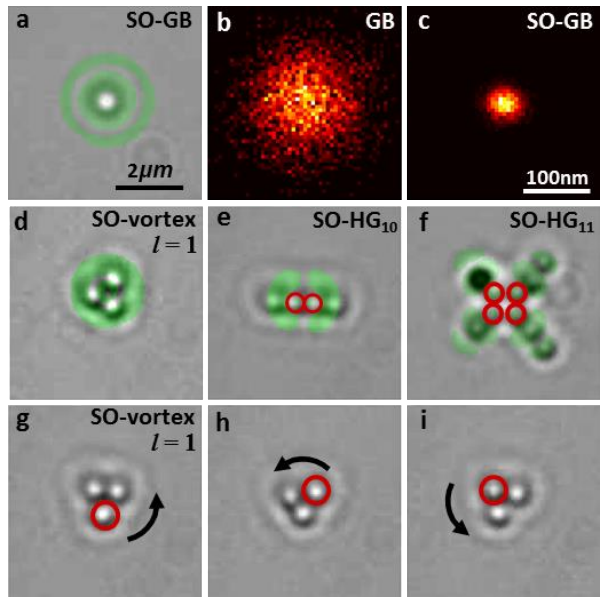

**Figure 3.** Particle manipulation with structured SO beams. **a** Trapping of single polystyrene bead of size $500 nm$ in diameter with the SO-Gaussian beam (SO-GB). **b** and **c** correspond to 2d probability distribution of position of a single trapped beads of the normal Gaussian beam (GB) and SO-Gaussian beam, respectively. Figs. **d-f** in second row show the multiple particle manipulations with the SO-vortex $l = 1$, SO-HG$_{10}$, and SO-HG$_{11}$ beams, respectively. (**g-i**) A time series showing the anticlockwise rotation of beads trapped in a SO-vortex beam of charge $l = 1$, due to transfer of orbital angular momentum $l\hbar$ from the SO-vortex beam to the trapped beads.

Next, we studied particle manipulation with structured SO beams. Specifically, we demonstrated the ability of SO-vortex beam to transfer its orbital angular momentum (OAM) $\pm l\hbar$ to polystyrene beads, enabling to simultaneously trap and rotate them at the vicinity of the dark core which size is $\approx (200 \pm 58) nm$, see Figs. 3 g-i. Additionally, clockwise and counter-clockwise particle rotation is achieved by changing the sign of the topological charge of the SO-vortex beam (supplemental movies 3, 4). Moreover, when we trapped particles in both the inner and outer rings we could distinguish between their rotations and infer the direction of the inner ring rotation. Also we observed that in the SO-vortex beam profile both the inner and outer rings have the same sign of topological charge. Trapping multiple particles in sub-diffraction limited traps can be done using SO-HG beams. In Fig. 3e (3f) we show two (four) particles, each one trapped in each one of the two (four) lobes of the SO-HG$_{10}$ (SO-HG$_{11}$) beam, each one having a lobe size $\approx (200 \pm 58) nm$. Higher order SO-HG beams are utilized for trapping of a single particle in each one of the beams' lobes, enabling to control the number of trapped particles in a tractable manner by changing the beam shape in a sequence from HG$_{22}$ to HG$_{10}$, see Supplemental movie 5. We observed that trapping occurs both in the ring and in the center at the same height as can be seen from the beam profile (Fig. 1b) and from an image of trapped particles which appear at the same height (see figure 6(b) in the Supplemental Material section). In other words, since we use high-index spheres the outer rings will attract our particles and therefore destabilize their trapping in the central trap rather than increase their trapping stiffness. At low laser powers we indeed see particles jumping from the center trap to the outer rings of the SO-GB. The same phenomena occur for the SO-vortex and SO-HG beams. Therefore, it is observed that all the generated SO beams, despite their lower intensity (but thanks to their smaller sub-diffraction size), have the ability to trap and manipulate particles.

To better understand the mechanism of the dramatic trapping enhancement in SO beams, we repeat the experiments with an IR laser of wavelength $\lambda = 1083 nm$. By using a longer wavelength we can tailor the width of the SO beam to approximate the size of the trapped particle (500nm), i.e., the measured FWHM of the central lobe of SO beam is $\approx (454 \pm 60) nm$ for $r_\pi = 0.65 r_{max}$, which minimizes the effect of the outer lobes of the SO beam on the trapped particle. In these conditions the FWHM of the normal beam is $\approx (808 \pm 60) nm$ which agrees well with the theoretical diffraction limit $\approx 0.38 \lambda / NA_{eff} = 748.25 nm$ with $NA_{eff} = 0.55$. The larger dimensions of the SO beam also facilitate its intensity profile characterization. For equal input laser power the peak intensity of central SO lobe is about 20 times less than that of the normal beam's peak intensity. For comparison purposes we maintain equal peak intensity for both the normal and SO beam traps, throughout the trapping experiments by tuning the input laser power (see Fig. 4a). To ensure trapping is in the focal plane of the SO beam we demonstrate simultaneous trapping both in the center and outer lobes of the SO beams (see figure 7 in the Supplemental Material section).

Further insight into the enhanced trapping of the SO beam is obtained modelling the interaction of the light beam with a single particle, by approximating the optical trapping force[37] at the experimental conditions by, $F_{grad} = \frac{\alpha}{2} \nabla I$ (Fig. 4b), and implementing it in a Brownian dynamics simulation of particle trapping[40], where α is the particle polarizability (see Supplemental section for full description and Fig. 5). In Fig. 4b the gradient force calculated theoretically from the known SO and Gaussian fields are compared, the maximal trapping force in the SO beam is nearly 2.3 times greater than that of normal beam. The enhanced trapping force results in an enhancement of trap stiffness by a factor of 9.1 experimentally (Fig. 4c, Supplemental movie 6), from $K_{GB} \approx 0.44 \, pN/\mu m$ for the Gaussian beam to $K_{SO} \approx 4.0 \, pN/\mu m$ for the SO beam, and a factor of 4.1 in simulations ($K_{GB} \approx 0.44 \, pN/\mu m$, $K_{SO} \approx 1.8 \, pN/\mu m$). We find that localization is enhanced in the SO beams compared to normal beams in all experiments and simulations. In terms of trapping efficiency, the SO beam shows more efficient trapping by the factor of 80 times than the normal Gaussian beam. The difference in the trapping efficiency between the IR and green lasers is due to the different size of the central SO lobe and power. However, experimentally the effect is more pronounced with respect to the theoretical prediction, probably because we use a simplified one-dimensional model that ignores the size effects of the trapped particle. A more sophisticated model that using Mie scattering theory to calculate the optical forces in the optical trapping[41] would be required to explain more accurately the trapping enhancement.

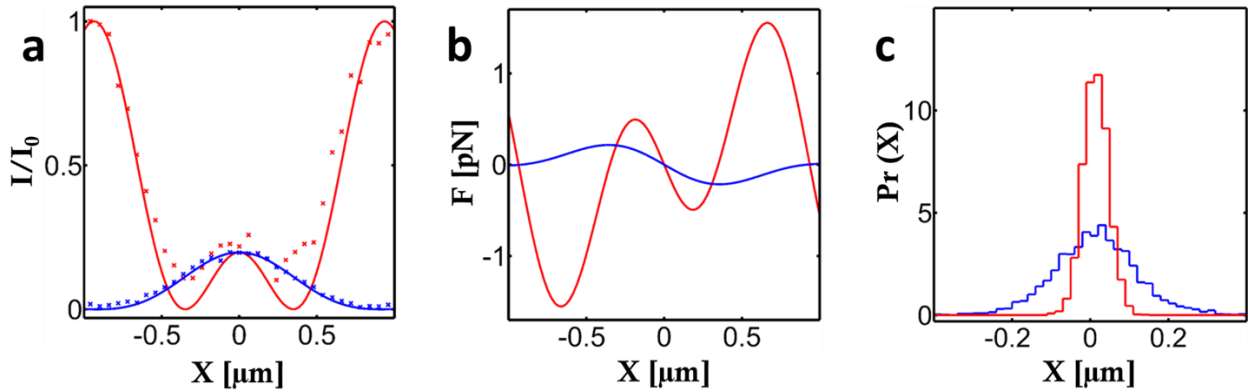

**Figure 4.** Trapping results with the IR laser. **a.** Normalized experimental (dots) and simulation (lines) 1d beam profile of the SO (red) and normal (blue) beams for equal peak intensity. **b** Computed gradient force of SO (red) and normal (blue) beams, where the vertical lines show the maximal gradient force of the central lobes of both the beams. **c.** Experimental 1d probability distribution of the single particle trapped in the central lobes of SO (red) and normal (blue) beams.

**CONCLUSIONS**

In conclusion, we have presented here a systematic method for the generation of structured super oscillatory beams focusing light beyond the standard diffraction limit. Further, we showed that these structured beams can confine nanometer polystyrene beads with unsurpassed precision and force. The localization accuracy is much higher than that achieved by diffraction limited Gaussian beam. The enhancement of the gradient force and the improved localization are also supported by a simple theoretical model, although the full mechanism responsible for the localization enhancement, merits more detailed theoretical analysis. The sub-diffraction spots of the structured beams may be applied for STED microscopy[11] (where SO-vortex beam and SO-Gaussian beam can be used for depletion and detection of fluorescent dyes), as well as in lithography. Finally, the method of structuring super-oscillating functions shown here can be used in other fields, e.g. nonlinear frequency conversion[31], plasmonics[42] as well as in the time domain for structuring light pulses for super-transmission[43] and for time-dependent focusing[44].

**ACKNOWLEDGEMENTS**

The authors would like to acknowledge the Israel Science Foundation (ISF) grant no. (1310/13) and Center for Nanoscience and Nanotechnology, Tel Aviv University for their financial support. We thank Yaniv Eliezer and Alon Bahabad for helpful discussion.